# TERAHERTZ BIOMEMS FOR ENZYMATIC CATALYSIS MONITORING

*Abdennour Abbas*[1,2,3], *Anthony Trezeibre*[1], *Nourredine Bourzgui, Didier Guillochon*[2], *Dominique Vercaigne-Marko*[2], *Philippe Supiot*[3] *and Bertrand Bocquet*[1]

[1] Institute of Electronics, Microelectronics and Nanotechnology, University of Lille1, France.
[2] Laboratory of Biological Process, Enzyme and Microbial Engineering, University of Lille1. FR
[3] Laboratory of Process Engineering of Reactive fluids-Materials Interactions, University of Lille1, FR

## ABSTRACT

This work focuses on the fabrication and use of a Biological Micro-ElectroMechanical System (BioMEMS) for TeraHertz (THz) detection and characterization of enzymatic catalysis reactions. The fluidic microdevice was fabricated using traditional lithographic techniques. It was then functionalized by reactive amine functions using plasma polymerized allylamine (pp-allylamine), followed by enzyme immobilization inside the microchannels. The enzymatic reaction was controlled by fluorescent microscopy before being analysed by sub-THz spectroscopy in the frequency range 0.06–0.11 THz. The preliminary results show the feasibility of terahertz monitoring of the biocatalysis reaction.

## INTRODUCTION

Today, there is an increasing interest in sensing biological samples with Terahertz (THz) spectroscopy. The interests of the THz spectrum arise from its potential to reveal collective vibrational modes related to the conformational movement of biomolecules in real time and without labelling [1]. Among the biological entities studied with THz radiations, the enzyme biocatalysis is one of the less explored topics. The enzyme activity can be investigated by THz since this activity is associated with conformational changes that occur during the conversion of substrates to products, such as protein hydrolysis into peptides after binding to a specific enzyme. However, the strong absorbance of water in the THz domain, make it necessary to work with small volumes. This becomes possible with the use of microsystems to draw microvolume of fluids until the THz sensor. Based on our previous work on the optimisation of THz transmission lines [2], and after have been demonstrated the validity of this approach on cell analysis [3], the current work aims to develop a Terahertz BioMEMS (Biological Micro-ElectroMechanical System) for real time monitoring of biocatalysis reactions.

## BIOMEMS FABRICATION

**Microfabrication**

To fabricate our THz BioMEMS, a mixed technology "quartz/polymer/silicon" was adopted. First, a single-wire transmission line, so called "Goubau Line" with a width of 5µm, was defined on a benzocyclobutene (BCB) coated quartz substrate. Secondly, the microchannels measuring 50µm width and 150µm depth were etched in a silicon wafer, which is subsequently bounded to the quartz substrate to have the transmission line under the microchannel (Fig. 1). The first level supporting both the microfluidic circuits and the THz probe was coated with a thin film to get primary amine functions on the surface. After the film deposition, the coated microchannels were covered with a BCB coated glass substrate to get buried fluidic circuits. Simulations performed with the designed structure show a good transmission of THz wave and low loss propagation (2.8 dB/mm at 0.14 THz).

**Aminofunctionalization**

As described above, the first level was functionalized before being covered by a glass substrate. This aminofunctionalization was performed by plasma polymerization of allylamine monomer in a home-built radio-frequency plasma reactor. A thin film polymer of approximately 100 nm was deposited with approximately 26 nm.s$^{-1}$ deposition rate. After thermal bounding of the two levels, connections of the bioMEMS to the surrounding environment are achieved by using microcapillaries.

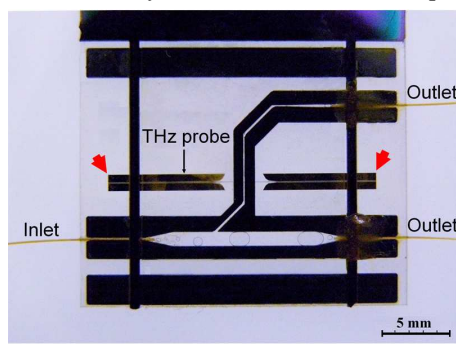

*Figure 1: Photography of the THz BioMEMS. The red arrows show the position of the vectorial network analyser tips.*

**Enzyme immobilization**

The last step of the bioMEMS fabrication is the enzyme immobilization on the microchannel walls. We chosed Trypsin (EC 3.4.21.4) as enzyme model. Using a microsyringe, a glutaraldehyde solution was injected into the microchannel and coupled to the aminated surface. This activation was followed by trypsin immobilization. The developed process allows us to obtain a surface covered with an enzyme monolayer. This feature is of a great importance in the analysis of enzymatic kinetics.

## ENZYME REACTION ANALYSIS
**Fluorescent microscopy analysis**

To demonstrate the efficiency of the immobilized enzymes, the bioactivity of trypsin was monitored by fluorescent microscopy. The specific substrate *Nα*-Benzoyl-L-Arginine-7-Amido-4-MethylCoumarin (BA-AMC) was injected into the functionalized channels, and its hydrolysis was detected according to the reaction:

BA-AMC + H$_2$O → *Nα*-Benzoyl-L-Arginine + 7-Amino-4-MethylCoumarin (AMC).

The fluorescent product shows that the immobilized enzymes remain active for multiple use cycles (Fig. 2).

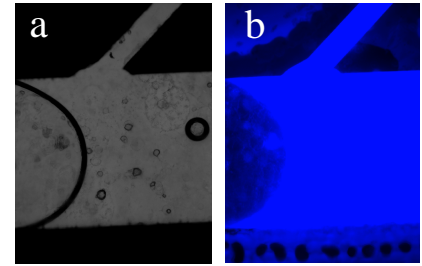

*Figure 2: The microchannel observed with fluorescence microscopy before (a) and after (b) the addition of the substrate BA-AMC. In blue, the fluorescent product (AMC) of the enzyme reaction.*

**Terahertz measurements**

The measurements were carried out using a vectorial network analyser, which emits and detects THz radiations at the frequency range 0-0.11 THz. Figure 3 show a good discrimination between liquids used for the measurements (water, PBS, Tris/HCl buffer, enzyme solution). A differentiation between the different solid materials used for the microchannel functionalization (pp-allylamine films, proteins) was also observed. This discrimination is achieved with a standard deviation of ± 0.07 dB. It is interesting to observe that the transmission changes show more clear and specific features over the frequency 60 GHz, which suggests that frequencies over this value are more suitable for transmission measurements. Furthermore, it can be seen in Figure 4 that the injection of BA-AMC into the functionalized microchannel leads to an abrupt and linear increase in transmission by approximately 1.28 ± 0.07 dB at 94 GHz. This reproducible change is presumably due to the linear increase of the enzyme activity. More investigations are needed to corroborate this statement.

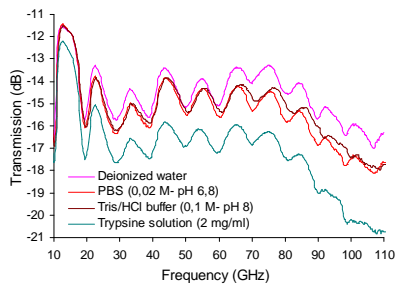

*Figure 3: THz transmission features of different liquids*

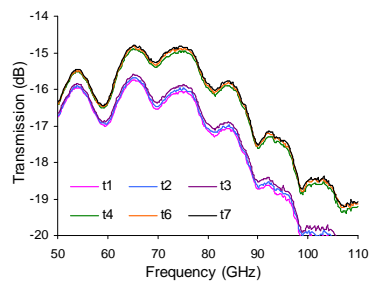 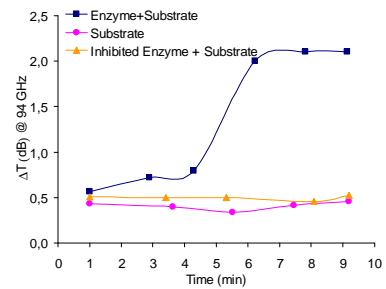

*Figure 4: THz transmission changes after addition of the substrate BA-AMC, "t" refers to time (min). The right graph reports the transmission variation (Δ T) as a function of time.*

## CONCLUSION
A bioMEMS was fabricated and functionalized by plasma polymerization followed by enzyme attachment. Fluorescence detection confirmed the efficiency of the enzyme reaction on the microchannel and sub-THz transmission analysis showed a good discrimination between the different liquids used here. Moreover, the preliminary analysis revealed that THz transmission intensity is sensitive the enzyme biocatalysis. Additional THz transmission experiments are in progress to define the changes in the transmission features of the enzyme reaction in respect to the substrate concentration. This will aid to determine the kinetic parameters of the reaction at the microscale level and enable the study of the microchannel effects on these parameters.